\documentclass[floatfix,prb,twocolumn]{revtex4}
\usepackage{graphicx}
\usepackage{psfrag}

\begin{document}

\title{Simultaneous readout of two charge qubits}

\author{Jian Li$^{1,2}$ and G{\"o}ran Johansson$^1$}

\affiliation{$^1$Microtechnology and Nanoscience, MC2, Chalmers,
S-412 96 G\"oteborg, Sweden \\
$^2$Nanoscience Center, P.O. Box 35, FIN-40014, University of
Jyv\"askyl\"a, Finland }

\begin{abstract}
We consider a system of two solid state charge qubits, coupled to a
single read-out device, consisting of a single-electron transistor
(SET). The conductance of each tunnel junction is influenced by its
neighboring qubit, and thus the current through the transistor is
determined by the qubits' state. The full counting statistics of the
electrons passing the transistor is calculated, and we discuss qubit
dephasing, as well as the quantum efficiency of the readout. The
current measurement is then compared to readout using real-time
detection of the SET island's charge state. For the latter method we
show that the quantum efficiency is always unity. Comparing the two
methods a simple geometrical interpretation of the quantum
efficiency of the current measurement appears. Finally, we note that
full quantum efficiency in some cases can be achieved measuring the
average charge of the SET island, in addition to the average
current.
\end{abstract}

\maketitle

\section{Introduction}
A solid state charge qubit, formed by a single electron trapped in a
double quantum dot (DQD), is an interesting system for studying
quantum coherence in the solid state
\cite{HayashiPRL2003,PettaPRL04,GormanPRL2005,FujisawaROP2006}.

The state of charge qubits can be read out by sensitive
electrometers, such as the radio-frequency single-electron
transistor (RF-SET)\cite{Schoelkopf98}. The straightforward scheme
is to couple the charge qubit capacitively to the small SET
island\cite{AassimePRL01}. The SET acts as a variable resistor,
dependent on the qubit state. The SET is included in an LC-tank
circuit, and the qubit state can be detected by measuring the tank
circuits dissipation. Theoretical estimates for the back-action on
the qubit indicate that single-shot qubit readout is possible
\cite{makhlin00,OurSETreadout}.

In principle the RF-SET can also be used to detect the charge of a
QD \cite{LuNature2003,FujisawaAPL04}. But a conceptually even
simpler method is to place the QD close to a quantum point contact
(QPC)\cite{FieldPRL93}. The charge of the QD will influence the
transparency of the QPC, and by applying a driving voltage across
the QPC, the charge can be determined through a current measurement.
This method has been successfully used to read out the charge of a
DQD \cite{ElzermanPRB03,PettaPRL04,JohnsonPRB05}.

Starting with Ref.~\onlinecite{gurvitz97}, there are numerous
theoretical works on the measurement induced dephasing rate
$\Gamma_\varphi$ of a charge qubit read out by a QPC, see e.g.
Ref.~\onlinecite{AverinSukhorukovPRL2005} and references therein.
The dephasing rate can be compared with the measurement time
$t_{ms}$, and from fundamental principles of quantum
measurement\cite{Braginsky}, the quantum efficiency
$\eta=(t_{ms}\Gamma_\varphi)^{-1}$ has an upper bound of unity,
implying that one cannot distinguish two states without destroying
the quantum coherence between them.

An interesting combination of the SET and QPC measurement techniques
was proposed by Tanamoto and Hu\cite{tanamoto05}, and is also the
focus of our present paper. The setup allows for reading out two
charge qubits using a single SET. The two qubits are positioned
close to the SET source and drain junction respectively (see
Fig.~\ref{SetupFig}). Thus the state of the left/right qubit will
influence the conductance of the left/right tunnel junction. By
applying a source-drain voltage and measuring the current, both
qubits can be read out. In Ref.~\onlinecite{tanamoto05} the focus
was made on the transient dynamics of qubit-detector system, as well
as the ensemble averaged current.

In this paper we concentrate on the single-shot readout properties,
by calculating the measurement times from the noise and full
counting statistics\cite{levitov96} of the charge transport. The
quantum efficiency is obtained by comparing with the measurement
induced dephasing rates. From the quantum inefficiency, in detecting
certain qubit states, we conclude that the qubits get entangled with
degrees of freedom not measured in current measurements. This leads
to an analysis of a measurement setup, where instead of the current,
the charge of the SET island is detected, in real-time. We find that
this measurement always give full quantum efficiency.

The outline of the paper is as follows. In Secs. \ref{ModelSection}
and \ref{KineticEquationSection} we derive a master equation for the
dynamics of the density matrix of the detector-qubit system, using
an open quantum systems approach\cite{GardinerZoller_Carmichael}. In
Sec.~\ref{ReadOutSection} we use this formulation to obtain the
single-shot measurement time, as well as  the dephasing rates, and
consequently also the quantum efficiency. We then apply the general
results to a few analytically tractable cases. In
Sec.~\ref{IslandStateMeasSec} the analysis is extended to the case
when the charge of the SET island is monitored. The properties of
that setup is compared with the current measurements in
Sec.~\ref{CurrentMeasurementsRevisitedSection}, leaving the
conclusion and discussion for Sec.~\ref{DiscussionSection}.

\section{The model}
\label{ModelSection} 
\begin{figure}[!ht]
\includegraphics[width=9cm]{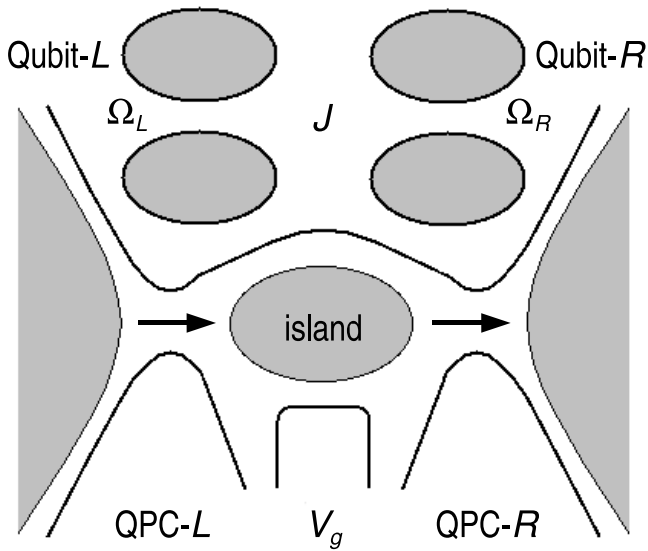}
\caption{A Coulomb blockade island in a single-electron transistor
geometry. The conductance of each tunnel junction is influenced by
a nearby charge qubit. (See text.)}\label{SetupFig}
\end{figure}

We consider two double-dot charge qubits, read
out by a quantum dot single-electron transistor. The SET island
contains one spin-degenerate level with energy $E_d$, and the
Coulomb interaction gives rise to an additional energy $U$ when it
is doubly occupied. Each qubit is electrostatically coupled to one
of the SET's quantum point contacts, as shown in Figure
\ref{SetupFig}.
There is also an electrostatic interaction between the qubits,
causing a direct qubit-qubit coupling with strength $J$. The two
qubits have a Hamiltonian of the following form
\begin{equation}
\label{eq5.1} H_{qb} = \sum_{\alpha=L,R} \big( \Omega_\alpha
\sigma_{\alpha x} + \Delta_\alpha \sigma_{\alpha z} \big) +
J\sigma_{Lz}\sigma_{Rz} ,
\end{equation}
where $\sigma_{\alpha x}$ and $\sigma_{\alpha z}$ are Pauli matrices
($\alpha\in\{L,R\}$). $\Omega_\alpha$ denotes the tunnel coupling
strength between the two dots of each qubit, and $\Delta_\alpha$ the
charging energy difference within each qubit. We can also express
the  Pauli matrices in terms of the creation and destruction
operators of the extra electron in the qubit quantum dots,
\begin{equation}
\label{eq5.2}
    \sigma_{\alpha x} = a_\alpha^\dag b_\alpha + b_\alpha^\dag a_\alpha ,
    \ \ \ \ \sigma_{\alpha z} = a_\alpha^\dag a_\alpha - b_\alpha^\dag b_\alpha ,
\end{equation}
where $a_\alpha^\dag$ and $a_\alpha$ denote the creation and
destruction operators in the upper dot of the $\alpha\in\{L,R\}$
qubit, and $b_\alpha^\dag$ and $b_\alpha$ operates on the lower dot.

The SET Hamiltonian can be split into three pieces: $H_{SET} =
H_{res} +  H_{is} + H_T $. $H_{res}$ describes the left and right
reservoirs
\begin{equation}
\label{eq5.3} H_{res} = \sum_{\alpha=L,R; s=\uparrow,\downarrow}
\sum_k E_{k\alpha} c_{k\alpha s}^\dagger c_{k\alpha s} ,
\end{equation}
where $E_{k\alpha}$ is the electron energy in the $\alpha$ reservoir
at wave vector $k$, and $c_{k\alpha s}^\dagger$ and $c_{k\alpha s}$
are corresponding electron creation and destruction operators. For
the SET island, we have,
\begin{equation}
\label{eq5.4} H_{is} = \sum_{s=\uparrow,\downarrow}E_d d_s^\dag
d_s + U d_{\uparrow}^\dag d_{\uparrow} d_{\downarrow}^\dag
d_{\downarrow} ,
\end{equation}
where $d_s^\dag$ and $d_s$ are creation and destruction operators of
the electrons on the island. We consider the transport through the
QPCs to be in the low transparency regime, described by a tunnel
Hamiltonian,
\begin{equation}
\label{eq5.5} H_T = \sum_{\alpha=L,R; s=\uparrow,\downarrow}
\sum_k V_{k\alpha} \left( c_{k\alpha s}^\dag d_s e^{i\chi_\alpha}
+ d_s^\dag c_{k\alpha s}e^{-i\chi_\alpha} \right) ,
\end{equation}
where we assume that the tunnelling strengths $V_{k\alpha}$ of
electrons between the reservoirs and the Coulomb island are
independent of the spin degree of freedom, and the spin $s$ is
conserved during the tunnelling processes. The counting fields
$\chi_\alpha$ for $\alpha\in\{L,R\}$ are introduced to keep track
of the number of electrons $m_{L/R}$ which passed through the
$L/R$ junction from the SET island to the $L/R$ lead. The
operators $e^{\pm i\chi_\alpha}$ changes this quantum number by
one, $e^{\pm i\chi_\alpha}|m_\alpha \rangle =|m_\alpha \pm 1
\rangle$.

The qubits electrostatically interact with the QPCs, and influence
the tunnelling rates according to the following Hamiltonian,
\begin{equation}
\label{eq5.6} H_{T}^{qb} = \sum_{\alpha,s,k} \delta V_{k\alpha}
\left( c_{k\alpha s}^\dag d_se^{i\chi_\alpha} + d_s^\dag
c_{k\alpha s}e^{-i\chi_\alpha} \right) \sigma_{\alpha z} ,
\end{equation}
where $\delta V_{k\alpha}$ are coupling strengths between the QPCs
and the qubits. Depending on the qubits' states, the tunnelling
strengths of the left and right QPCs change from $V_{k\alpha}$ to
$V_{k\alpha} \pm \delta V_{k\alpha}$. For simplicity, we neglect
the relative phase between these tunnelling strengths, but we
still allow for a weak energy dependence of their magnitudes.

\section{A kinetic equation}
\label{KineticEquationSection} One conventional approach to
continuous quantum measurement problems is called the open system
approach \cite{GardinerZoller_Carmichael}. In general terms, an open
system is a limited quantum system coupled to another quantum system
with a large number of degrees of freedom, called the environment.
In our case, the reservoirs (leads) connected to the left and right
QPCs constitute an environment for the limited quantum system
consisting of the two qubits and the SET island. The system
hamiltonian is $H_S=H_{qb}+H_{is}$, while the hamiltonian of the
environment is $H_{res}$ and the interaction hamiltonian is given by
the tunnelling terms $H_{int}=H_T+H_{T}^{qb}$.
\begin{widetext}
The starting point is the Liouville equation on
integro-differential form
\begin{equation}
\label{A7}
    \partial_t \tilde{\rho}_{tot}(t) = -i [ \tilde{H}_{int}(t),\rho_{tot} (0) ] -
    \int_0^t dt' \left[ \tilde{H}_{int}(t), [ \tilde{H}_{int}(t'),\tilde{\rho}_{tot}(t') ] \right]
    ,
\end{equation}
where $\tilde{H}_{int}(t)$ is the interaction hamiltonian and
$\tilde{\rho}_{tot}(t)$ the density matrix in the interaction
picture. Then we proceed to a markoffian kinetic equation for the
reduced density matrix of the system $\tilde{\rho}(t)$, making the
Born and Markov approximations,
\begin{equation}
\label{A10} \dot{\tilde{\rho}}(t) = - \int_0^t dt'
{\mathrm{Tr}}_{res} \left\{ \left[ \tilde{H}_{int}(t),[
\tilde{H}_{int}(t'),\tilde{\rho}(t) ] \right] \right\} ,
\end{equation}
where ${\mathrm{Tr}}_{res}$ indicates taking the trace over the
reservoir degrees of freedom, which are taken to be in thermal
equilibrium, at their respective chemical potential. The integrals
on the right-hand side of Eq.~(\ref{A10}) can be approximately
evaluated, taking the lower limit of the integral to minus infinity.
They have in general both real and imaginary parts, where the
imaginary parts effectively can be absorbed into renormalized system
energies. The real parts give the dissipative dynamics, i.e. the
electron tunnelling rates. A typical example of a term contributing
to the tunnelling rate from lead $\alpha$ to the island is
\begin{equation}
\int_0^\infty d\tau \sum_{k,s} e^{\pm i(E_{ks}-E_d)\tau}
V_{k\alpha}^2 f_\alpha(E,T) \rightarrow \pi g_\alpha(E_d)
V_\alpha^2(E_d) f_\alpha(E_d,T) ,
\end{equation}
where we have taken the sum to an integral by introducing the
density of states in the leads $g_\alpha(E)$, and the energy
dependent tunnelling amplitude $V_\alpha(E)$, and $f_\alpha(E,T)$
is the Fermi function of lead $\alpha$. The bias voltage between
the left and right lead enters through the shifted Fermi
distributions. For details, please see
Ref.~\onlinecite{Li06diplom}.

\subsection{The non-zero tunnelling rates}
Since our quantum system has 16 degrees of freedom the reduced
density matrix has 256 elements, and thus the kinetic equation in
general 256 $\times$ 256 terms, we have to make some simplifying
assumptions to arrive at a tractable set of equations.

First we consider a source-drain voltage across the SET large
enough compared to temperature so that we may neglect backwards
tunnelling. Thus, without the coupling to the qubits, and with
spin-independent tunnelling, there are four different rates to
consider; tunnelling onto the unoccupied SET island from the left
lead ($\Gamma_L$), tunnelling onto the singly occupied SET island
from the left lead ($\Gamma'_L$), tunnelling off the singly
occupied SET island to the right lead ($\Gamma_R$), and finally
 tunnelling off the doubly occupied
SET island to the right lead ($\Gamma'_R$), see Fig.~\ref{RatesFig}.
\begin{figure}[!ht]
\includegraphics[width=9cm]{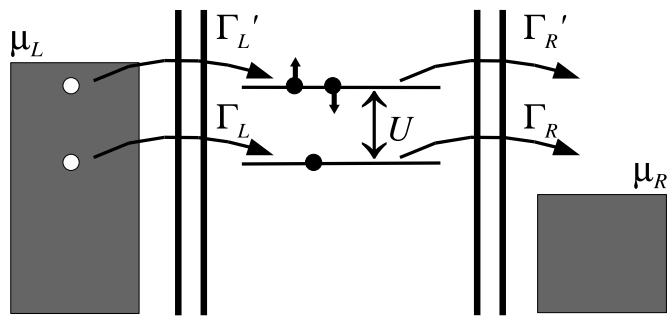}
\caption{The four non-zero tunneling rates for electrons, on and off
the SET island. The applied source-drain voltage $eV$ creates a
difference in chemical potentials between the left and right lead
$eV=\mu_L-\mu_R$.}\label{RatesFig}
\end{figure}

Including the coupling to the qubits each rate will acquire a
state-dependent shift $\Gamma_\alpha^\pm=\Gamma_\alpha\pm \delta
\Gamma_\alpha$ and $\Gamma^{'\pm}_\alpha=\Gamma'_\alpha\pm \delta
\Gamma'_\alpha$, see Fig.~\ref{QubitRatesFig}. Not to complicate the
system further we assume that qubit energies $\Delta_\alpha,
\Omega_\alpha$, and $J$ are small enough so that the tunnelling
rates including a flip of the qubits's state are identical to the
ones without flipping the qubit state.
\begin{figure}[!ht]
\includegraphics[width=6cm]{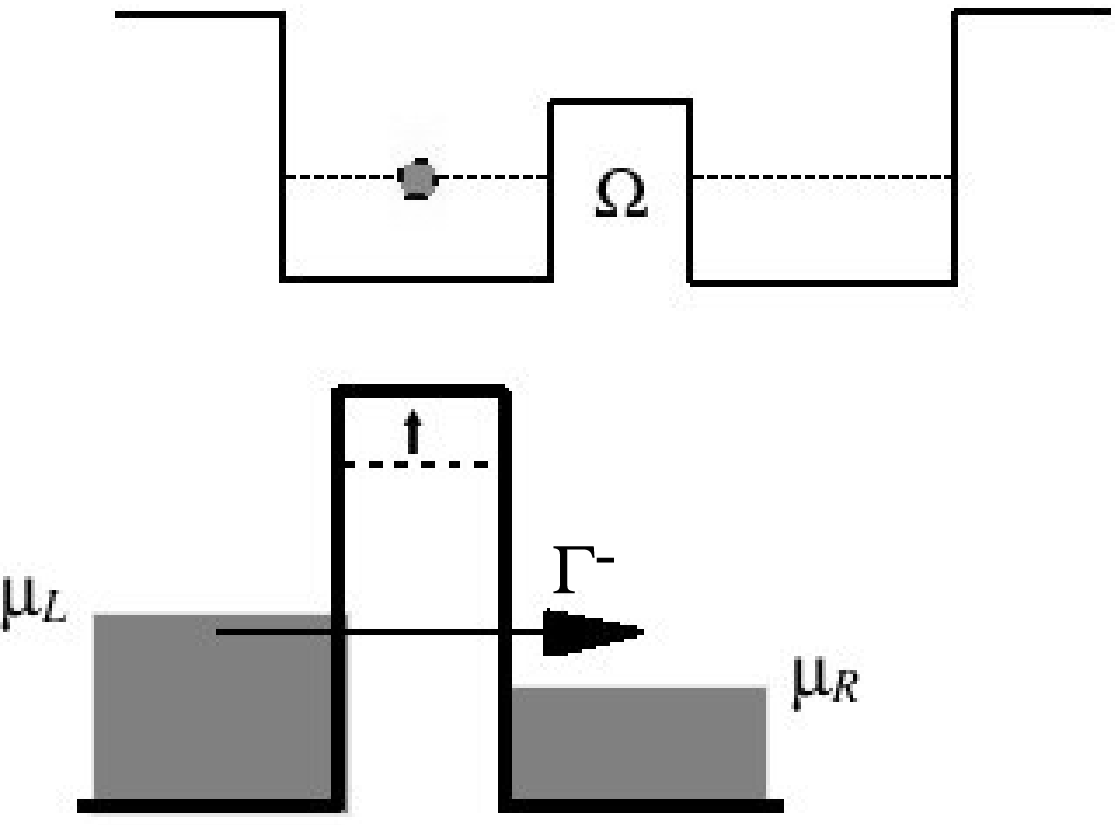} \ \ \
\includegraphics[width=6cm]{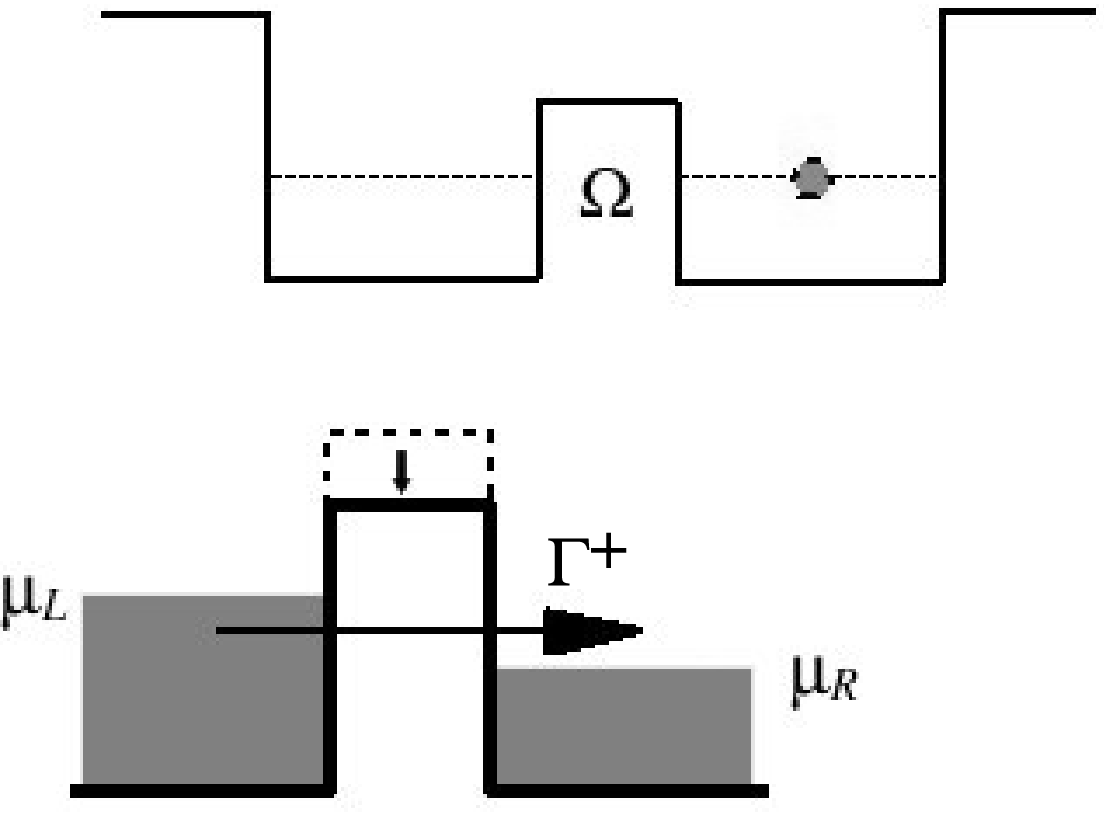}
\caption{An illustration of the capacitative coupling between a
double quantum dot qubit and low transparency point contact. Left:
When the electron is located in the dot closest to the PC, the
barrier is higher, leading to a lower tunnelling rate
$\Gamma^-=\Gamma-\delta\Gamma$. Right: For the qubit in the other
state, the barrier is lower and the tunnelling rate higher
$\Gamma^+=\Gamma+\delta\Gamma$. }\label{QubitRatesFig}
\end{figure}

We are now ready to present the kinetic equation for the elements of
the reduced density matrix. As basis we chose the four qubit product
states $|A\rangle = |\downarrow\downarrow\rangle, |B\rangle =
|\downarrow\uparrow\rangle, |C\rangle = |\uparrow\downarrow\rangle,
|D\rangle = |\uparrow\uparrow\rangle$, the SET island states
corresponding to no electron $|a\rangle$, one spin-up/down electron
$|b_\uparrow\rangle/|b_\downarrow\rangle$ and doubly occupied
$|c\rangle$, and the number of electrons which has tunnelled across
the right junction $|m\rangle$:
\begin{equation}
\label{B1}
    \rho_{z_1z_2}^u(m) = \langle m| \otimes \langle u | \otimes \langle z_1 | \ \rho \ | z_2
    \rangle \otimes | u \rangle \otimes |m\rangle ,
\end{equation}
where $z_1, z_2 \in \{A, B, C, D\}$ and $u \in \{a, b_\uparrow,
b_\downarrow, c\}$. Here we only keep terms off-diagonal in the
qubit states. It is easy to verify that terms off-diagonal in the
island state do not couple to the diagonal terms, and they decay on
a time-scale given by $\Gamma_{L/R}^{-1}$. Because the kinetic
equation is translationally invariant in $m$-space, we Fourier
transform it with respect to this variable $\rho_{z_1 z_2}^u(k)
\equiv \sum_m e^{ikm}\rho_{z_1z_2}^u(m)$
\cite{makhlin00,goan03,clerk03a}, and arrive at the following
kinetic equation for the elements of the reduced density matrix
\begin{eqnarray}
    \dot{\rho}_{z_1z_2}^a &=& [ i(J_{z_2} - J_{z_1}) -
(\Gamma_L^{z_1} + \Gamma_L^{z_2}) ]\rho_{z_1z_2}^a -
i\Omega_R\left[\rho_{g_r(z_1),z_2}^a -
\rho_{z_1,g_r(z_2)}^a\right] \nonumber \\
    &&- i\Omega_L\left[\rho_{g_l(z_1),z_2}^a -
\rho_{z_1,g_l(z_2)}^a\right] +
\sqrt{\Gamma_R^{z_1}\Gamma_R^{z_2}}e^{ik}\left(\rho_{z_1z_2}^{b_\uparrow}
+ \rho_{z_1z_2}^{b_\downarrow}\right) , \label{eq6.3} \\
    \dot{\rho}_{z_1z_2}^{b_\uparrow}  &=& \left[ i(J_{z_2} - J_{z_1})
    - \frac{\Gamma_L^{'z_1}+\Gamma_L^{'z_2}+\Gamma_R^{z_1}+\Gamma_R^{z_2}}{2}
    \right]\rho_{z_1z_2}^{b_\uparrow} - i\Omega_R\left[\rho_{g_r(z_1),z_2}^{b_\uparrow} -
\rho_{z_1,g_r(z_2)}^{b_\uparrow}\right] \nonumber \\
    &&- i\Omega_L\left[\rho_{g_l(z_1),z_2}^{b_\uparrow} -
\rho_{z_1,g_l(z_2)}^{b_\uparrow}\right] +
\sqrt{\Gamma_L^{z_1}\Gamma_L^{z_2}}\rho_{z_1z_2}^a +
\sqrt{\Gamma_R^{'z_1}\Gamma_R^{'z_2}}e^{ik}\rho_{z_1z_2}^c , \label{eq6.4} \\
    \dot{\rho}_{z_1z_2}^{b_\downarrow}  &=& \left[ i(J_{z_2} - J_{z_1})
    - \frac{\Gamma_L^{'z_1}+\Gamma_L^{'z_2}+\Gamma_R^{z_1}+\Gamma_R^{z_2}}{2}
    \right]\rho_{z_1z_2}^{b_\downarrow} - i\Omega_R\left[\rho_{g_r(z_1),z_2}^{b_\downarrow} -
\rho_{z_1,g_r(z_2)}^{b_\downarrow}\right] \nonumber \\
    &&- i\Omega_L\left[\rho_{g_l(z_1),z_2}^{b_\downarrow} -
\rho_{z_1,g_l(z_2)}^{b_\downarrow}\right] +
\sqrt{\Gamma_L^{z_1}\Gamma_L^{z_2}}\rho_{z_1z_2}^a +
\sqrt{\Gamma_R^{'z_1}\Gamma_R^{'z_2}}e^{ik}\rho_{z_1z_2}^c , \label{eq6.5} \\
    \dot{\rho}_{z_1z_2}^c  &=& [ i(J_{z_2} - J_{z_1}) -
(\Gamma_R^{'z_1} + \Gamma_R^{'z_2}) ]\rho_{z_1z_2}^c -
i\Omega_R\left[\rho_{g_r(z_1),z_2}^c -
\rho_{z_1,g_r(z_2)}^c\right] \nonumber \\
    &&- i\Omega_L\left[\rho_{g_l(z_1),z_2}^c -
\rho_{z_1,g_l(z_2)}^c\right] +
\sqrt{\Gamma_L^{'z_1}\Gamma_L^{'z_2}}\left(\rho_{z_1z_2}^{b_\uparrow}
+ \rho_{z_1z_2}^{b_\downarrow}\right) , \label{eq6.6}
\end{eqnarray}
where
\begin{eqnarray}
    \Gamma_L^A &=& \Gamma_L^B = \Gamma_L^- , \ \ \ \ \Gamma_L^C =
    \Gamma_L^D = \Gamma_L^+ , \ \ \ \ \Gamma_R^A = \Gamma_R^C = \Gamma_R^- , \ \ \ \ \Gamma_R^B =
    \Gamma_R^D = \Gamma_R^+ , \nonumber \\
    \nonumber \\
    \Gamma_L^{'A} &=& \Gamma_L^{'B} = \Gamma_L^{'-} , \ \ \ \ \Gamma_L^{'C} =
    \Gamma_L^{'D} = \Gamma_L^{'+} , \ \ \ \ \Gamma_R^{'A} = \Gamma_R^{'C} = \Gamma_R^{'-} , \ \ \ \ \Gamma_R^{'B} =
    \Gamma_R^{'D} = \Gamma_R^{'+} , \nonumber  \\
    \nonumber \nonumber \\
    J_A &=& - \Delta_L - \Delta_R + J , \ \ \ \ J_B = - \Delta_L +
    \Delta_R - J ,  \ \ \ \ J_C = \Delta_L - \Delta_R - J , \ \ \ \ \ \ J_D = \Delta_L +
    \Delta_R + J , \label{B7}
\end{eqnarray}
and
\begin{equation}
    g_l(A) = C , \ \ g_l(B) = D , \ \ g_l(C) = A , \ \ g_l(D) = B, \ \ g_r(A) = B , \ \ g_r(B) = A , \ \ g_r(C) = D , \ \ g_r(D) = C .
\label{B8}
\end{equation}
We note that, neglecting the counting field ($k \rightarrow 0$), the
above equations coincide with the rate equations derived by Tanamoto
and Hu (see Appendix in Ref. \onlinecite{tanamoto05}), using a
different approach\cite{gurvitz97}. Eqs.~(\ref{eq6.3}) to (\ref{B8})
form one of the main results of this paper, and they constitute the
basis of our further analysis. One may note that although the rates
$\Gamma_{L/R}^{(')}$ and $\delta \Gamma_{L/R}^{(')}$ are intimately
connected to the microscopic tunnelling hamiltonians in
Eqs.~(\ref{eq5.5}) and (\ref{eq5.6}), we may now consider different
operating regimes in terms of the value of these rates themselves,
rather than specifying the coefficients of the tunnel hamiltonians.

\subsection{Numerical methods}
To solve for the time-dependence of the $k$-dependent reduced
density matrix, we rewrite the kinetic equations on matrix form
$\dot{\vec{\rho}} = M\cdot\vec{\rho}$ as follows
\begin{equation}
\label{eq4.51}
    \left (
        \begin{array}{c}
        \dot{\rho}_{AA}^a \\
        \dot{\rho}_{AB}^a \\
        \vdots \\
        \dot{\rho}_{DD}^c
        \end{array}
    \right )_{64} =
    \left (
        \begin{array}{cccc}
        -2\Gamma_L^- & i\Omega_R & \cdots & 0 \\
        i\Omega_R & 2i(\Delta_R-J)-2\Gamma_L^- & \cdots & 0 \\
        \vdots & \vdots & \ddots & \vdots \\
        0 & 0 & \cdots & -2\Gamma_R^{'+}
        \end{array}
    \right )_{64 \times 64} \cdot
    \left (
        \begin{array}{c}
        \rho_{AA}^a \\
        \rho_{AB}^a \\
        \vdots \\
        \rho_{DD}^c
        \end{array}
    \right )_{64} .
\end{equation}
Thus the reduced density matrix elements at arbitrary time $t$ are
given by the evolution matrix $M$
\begin{equation}
\label{eq4.52}
    \vec{\rho}(t) = e^{Mt} \vec{\rho}(0) .
\end{equation}
By considering the time-derivative of the average number of
electrons which have tunnelled across the right junction $\langle
\dot{m} \rangle = Tr \{ m |m\rangle\langle m| \dot{\rho} \} = Tr \{
-i\partial_k \dot{\rho}(k=0) \} $ we get the expression for the
time-dependent average current
\begin{eqnarray}
\label{D7}
    \langle I_R(t) \rangle &=& e\sum_{z}\left\{ \Gamma_R^z\left[ \rho_{zz}^{b_\uparrow}(t) + \rho_{zz}^{b_\downarrow}(t)\right] +
    2\Gamma_R^{'z}\rho_{zz}^c(t) \right\},
    \label{averagecurrentformula}
\end{eqnarray}
which also agrees with the one used in Ref. \onlinecite{tanamoto05}.
To discuss the single-shot measurement time, needed to separate
different qubit states, we will also need the time-dependent
probability distribution of the number of electrons transferred
through the detector $P(m,t)$. This we obtain by tracing out the
qubits' and island's degrees of freedom,
\begin{equation}
\label{eq6.7}
    P(m,t)\equiv\sum_{z,u}\rho(z,z;u,u;m,m;t) ,
\end{equation}
which is straightforward using the Fourier transformed
$k$-dependent density matrix,
\begin{equation}
\label{eq6.8}
    P(m,t) = \int_{-\pi}^\pi\frac{dk}{2\pi}e^{-ikm}P(k,t) =
    \int_{-\pi}^\pi\frac{dk}{2\pi}e^{-ikm}\sum_{z,u}\rho_{z,z}^u(k,t) .
\end{equation}
We are now ready to discuss the properties of reading out two charge
qubits using a single-electron transistor, in a few different
parameter regimes.

\section{Read-out in the qubits' eigenbasis}
\label{ReadOutSection} To approach a quantum non-demolition (QND)
measurement for a weakly coupled read-out device, it is necessary to
measure in the qubits' eigenbasis. For the setup considered here,
the measurement is performed in the $z$-direction of each qubit,
i.e. the measurement determines in which dot the electron of each
qubit is located. Thus we consider the case when the inter-dot
coupling is switched off, i.e. $\Delta_L=\Delta_R=0$. We then start
from Eqs.~(\ref{eq6.3}-\ref{B8}) and derive simplified kinetic
equations. Due to the spin degeneracy of the tunneling, we may treat
the two singly occupied island states together by defining
$\rho_{z_1z_2}^{b}=\rho_{z_1z_2}^{b_\uparrow}+\rho_{z_1z_2}^{b_\downarrow}$.
Also, we note that the terms proportional to $i(J_{z_2}-J_{z_1})$
vanish by going to the rotating frame, with respect to these
coherent $z$-rotations, i.e. $\tilde{\rho}_{z_1z_2}(t) =
e^{-i(J_{z_2}-J_{z_1})t} \rho_{z_1z_2}(t)$. We arrive at the
following equations for the density matrix in the rotating frame
$\tilde{\rho}(t)$,
\begin{eqnarray}
    \dot{\rho}_{z_1z_2}^a &=& -
(\Gamma_L^{z_1} + \Gamma_L^{z_2})\rho_{z_1z_2}^a +
\sqrt{\Gamma_R^{z_1}\Gamma_R^{z_2}}\rho_{z_1z_2}^{b} , \\
    \dot{\rho}_{z_1z_2}^{b} &=& - \frac{\Gamma_L^{'z_1}+\Gamma_L^{'z_2}+\Gamma_R^{z_1}+\Gamma_R^{z_2}}{2}
    \rho_{z_1z_2}^{b} + 2\sqrt{\Gamma_L^{z_1}\Gamma_L^{z_2}}\rho_{z_1z_2}^a +
2\sqrt{\Gamma_R^{'z_1}\Gamma_R^{'z_2}}\rho_{z_1z_2}^c , \\
    \dot{\rho}_{z_1z_2}^c &=& - (\Gamma_R^{'z_1} + \Gamma_R^{'z_2})\rho_{z_1z_2}^c +
\sqrt{\Gamma_L^{'z_1}\Gamma_L^{'z_2}}\rho_{z_1z_2}^{b} ,
\end{eqnarray}
where we have suppressed the 'tilde' for notational brevity.

\subsection{Dephasing rates and the quasistationary distribution}
Since the interdot coupling is switched off ($\Delta_L=\Delta_R=0$),
the equations for density matrix elements with different qubit state
indices ($z_1z_2$) separate. To analyze these equations further we
now consider the elements of the reduced qubit density matrix
$\rho_{z_1z_2}=\rho_{z_1z_2}^a+\rho_{z_1z_2}^b+\rho_{z_1z_2}^c$,
obeying the equation
\begin{equation}
\dot{\rho}_{z_1z_2}=-\Gamma_{\varphi L}^{z_1z_2} \rho^a_{z_1z_2} -
\frac{1}{2} (\Gamma_{\varphi R}^{z_1z_2}+\Gamma_{\varphi
L}^{'z_1z_2}) \rho^b_{z_1z_2}- \Gamma_{\varphi R}^{'z_1z_2}
\rho^c_{z_1z_2}, \label{dephasing1}
\end{equation}
where we defined the basic dephasing rates
\begin{equation}
\Gamma_{\varphi L/R}^{z_1z_2}=
\left(\sqrt{\Gamma_{L/R}^{z_1}}-\sqrt{\Gamma_{L/R}^{z_2}}\right)^2 \
{\rm and } \ \ \Gamma_{\varphi L/R}^{'z_1z_2}=
\left(\sqrt{\Gamma_{L/R}^{'z_1}}-\sqrt{\Gamma_{L/R}^{'z_2}}\right)^2.
\label{DephasingRates}
\end{equation}
It is clear that these rates are only non-zero for off-diagonal
elements ($z_1 \neq z_2$). Thus the diagonal elements $\rho_{zz}$
are conserved, i.e. $\dot{\rho_{zz}}=0$, as required by probability
conservation. However, all off-diagonal elements decay, which is the
definition of dephasing. From this we may conclude that the
measurement will eventually destroy all coherence between the
different qubit product states $|\downarrow\downarrow\rangle,
|\downarrow\uparrow\rangle, |\uparrow\downarrow\rangle$ and
$|\uparrow\uparrow\rangle$. The expressions in
Eq.~(\ref{DephasingRates}) are similar to the ones presented in
Ref.~\onlinecite{tanamoto05}, and for each junction and tunnel event
separately the rates agrees with the result for a single qubit read
out by a single QPC.\cite{gurvitz97}

In the regime of weak measurement $\delta\Gamma_{L/R}/\Gamma_{L/R}
\ll 1$, the dephasing rates are small
\begin{equation}
\Gamma_{\varphi} = \frac{\delta\Gamma^2}{\Gamma} \left[ 1 +
O\left(\frac{\delta\Gamma^2}{\Gamma^2}\right) \right] ,
\end{equation}
so the dynamics will have two different timescales. On the short
timescale given by $\Gamma_{L/R}^{-1}$, all reduced density matrix
elements will be approximately conserved, and we obtain the
quasistationary distribution
\begin{eqnarray}
\frac{\rho_{z_1z_2}^{a}(t)}{\rho_{z_1z_2}(t)}&=&\frac{\Gamma_R^{z_1z_2}\Gamma_R^{'z_1z_2}}
{\Gamma_L^{z_1z_2}\Gamma_L^{'z_1z_2}+2\Gamma_L^{z_1z_2}\Gamma_R^{'z_1z_2}
+
\Gamma_R^{z_1z_2}\Gamma_R^{'z_1z_2}} \equiv P^a_{z_1z_2}, \nonumber\\
\frac{\rho_{z_1z_2}^{b}(t)}{\rho_{z_1z_2}(t)}&=&\frac{2\Gamma_L^{z_1z_2}\Gamma_R^{'z_1z_2}}
{\Gamma_L^{z_1z_2}\Gamma_L^{'z_1z_2}+2\Gamma_L^{z_1z_2}\Gamma_R^{'z_1z_2}+
\Gamma_R^{z_1z_2}\Gamma_R^{'z_1z_2}} \equiv P^b_{z_1z_2}, \nonumber\\
\frac{\rho_{z_1z_2}^{c}(t)}{\rho_{z_1z_2}(t)}&=&\frac{\Gamma_L^{z_1z_2}\Gamma_L^{'z_1z_2}}
{\Gamma_L^{z_1z_2}\Gamma_L^{'z_1z_2}+2\Gamma_L^{z_1z_2}\Gamma_R^{'z_1z_2}+
\Gamma_R^{z_1z_2}\Gamma_R^{'z_1z_2}} \equiv P^c_{z_1z_2},
\label{quasistationarydist}
\end{eqnarray}
where $\Gamma_{L/R}^{z_1z_2}=\Gamma_{L/R}^{z_1}+\Gamma_{L/R}^{z_2}$,
 $\Gamma_{L/R}^{'z_1z_2}=\Gamma_{L/R}^{'z_1}+\Gamma_{L/R}^{'z_2}$, and
 $P_{zz}^\alpha$ is the probability to find the island in the state $\alpha \in \{a,b,c\}$
 given that the qubits are in the product state $|z\rangle \in \{|A\rangle, |B\rangle,|C\rangle,|D\rangle\}$.
In order to get a qualitative understanding it is useful to note
that According to Eq.~(\ref{dephasing1}), the dephasing will take
place on the much longer timescale
$\Gamma_{L/R}/\delta\Gamma_{L/R}^{2}$, and as we will see, this is
also the time-scale for the measurement. Inserting this
quasi-stationary distribution into Eq.~(\ref{dephasing1}) we find
the total dephasing rate between the product states $|z_1\rangle$
and $|z_2\rangle$,
\begin{equation}
\Gamma_\varphi^{z_1z_2}=\frac{1}{2}\left[ P^a_{z_1z_2} 2
\Gamma_{\varphi L}^{z_1z_2} + P^b_{z_1z_2}\left(\Gamma_{\varphi
R}^{z_1z_2} + \Gamma_{\varphi L}^{'z_1z_2}\right)+P^c_{z_1z_2} 2
\Gamma_{\varphi R}^{'z_1z_2} \right]. \label{TotalDephasingRate}
\end{equation}
One simple interpretation of this formula is that the coherence
between state $z_1$ and $z_2$ will decrease only due to tunnel
events across a junction where the two states differ.

\subsection{Average current, noise and the measurement time}
The average current will approach a steady-state value on the short
time-scale given by $\Gamma_{L/R}^{-1}$ (see
Fig.~\ref{AverageCurrentFig}). In the case when the qubits initially
are in the product state $|z\rangle$ the steady-state average
current
\begin{equation}
I_{z}=e \left[P^b_{zz} \Gamma_R^{z} + P^c_{zz} 2
\Gamma_R^{'z}\right],
\end{equation}
is obtained by setting $\rho_{zz}(t)=1$ in
Eq.~(\ref{quasistationarydist}) and then inserting $\rho_{zz}^a,
\rho_{zz}^b=\rho_{zz}^{b\uparrow}+\rho_{zz}^{b\downarrow}$ and
$\rho_{zz}^c$ in Eq.~(\ref{averagecurrentformula}).
\begin{figure}
    \includegraphics[width=9cm]{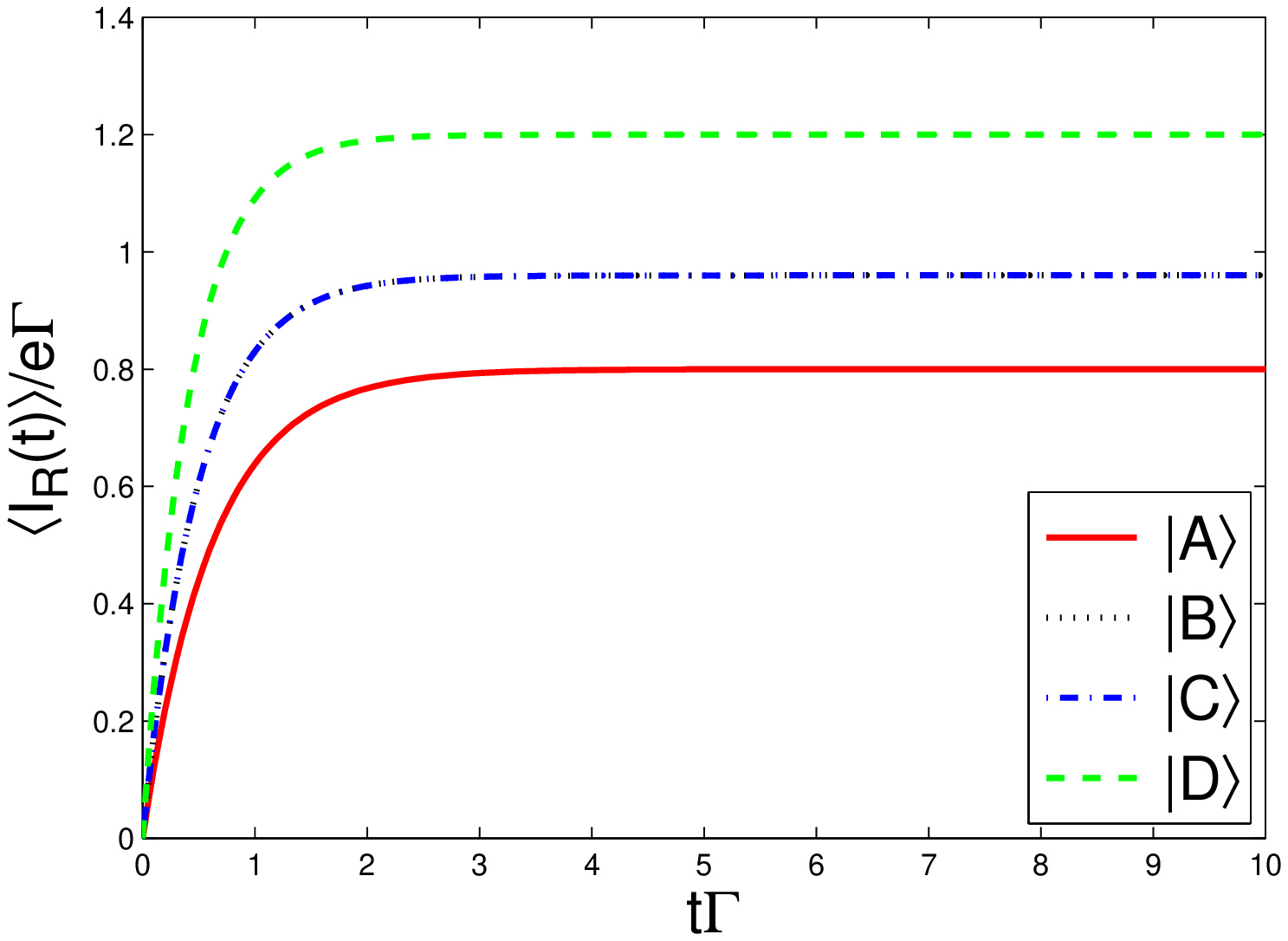}\includegraphics[width=9cm]{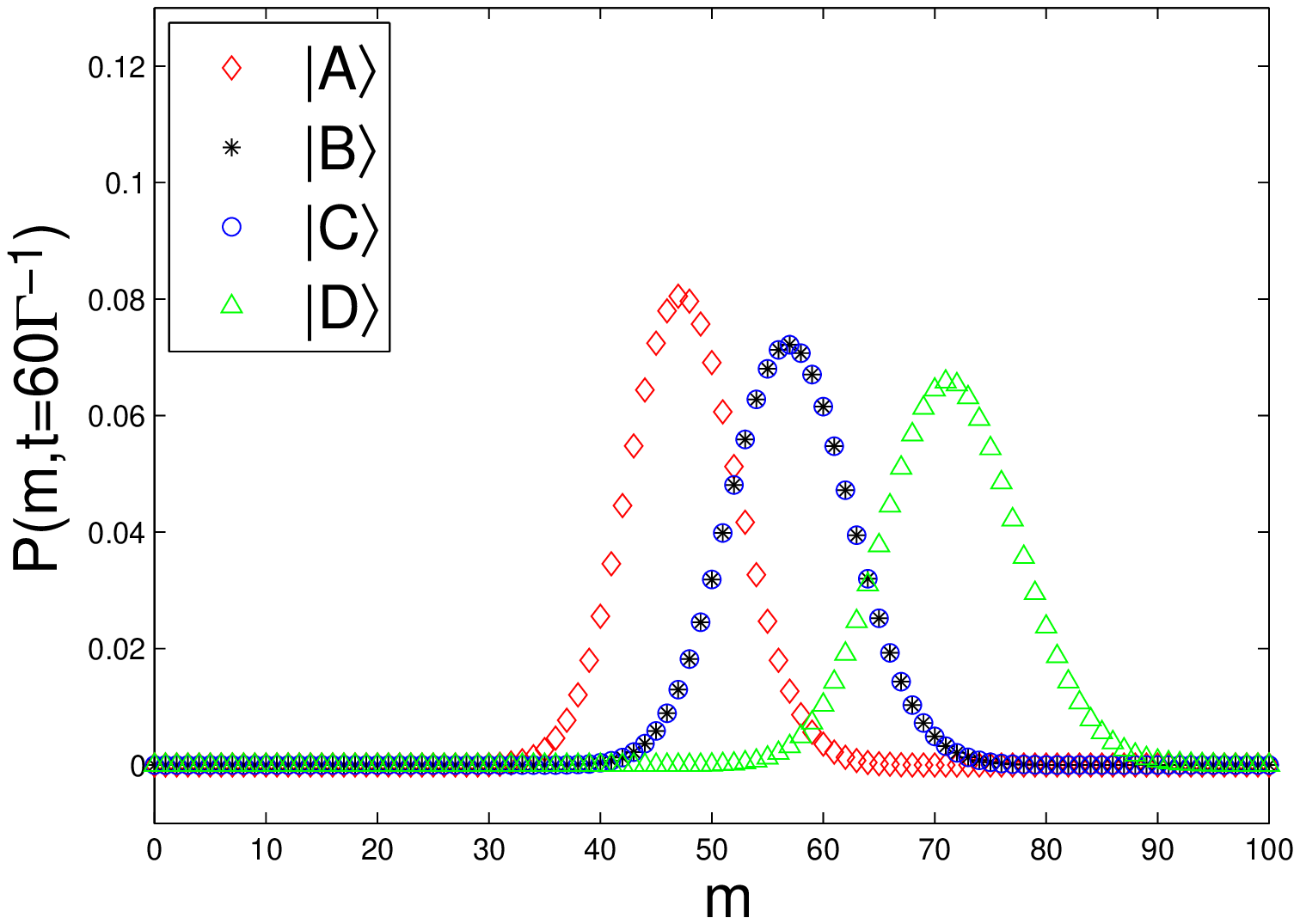}
    \caption{Left: The average current through the SET for the different initial qubit product
    states, with no qubit interdot coupling ($\Delta = 0$), and $\delta\Gamma_L = \delta\Gamma_R = 0.2 \Gamma$.
    Right: The corresponding probability distributions $P(m,t)$ for the number of electrons $m$ having passed the SET at time $t=60/\Gamma$.}
    \label{AverageCurrentFig}
\end{figure}

For times much longer than $\Gamma_{L/R}^{-1}$, when the average
number of electrons which has tunnelled through the SET is large
$\langle m \rangle \gg 1$, the distribution $P(m,t)$ will approach a
Gaussian with average $\langle m \rangle \approx I_z t /e$, and a
width given by the zero-frequency current noise $S_I$. The current
noise is readily obtained from the full counting statistics as
\cite{clerk03a,BagretsPRB03}
\begin{equation}
S_I=2e^2\left(-\frac{\partial^2}{\partial k^2} \lambda_0(k)\right),
\end{equation}
where $\lambda_0(k)$ is the eigenvalue of the evolution matrix $M$
which approaches zero in the limit $k\rightarrow 0$. The noise is
sometimes expressed in terms of the Fano factor $f$, through the
relation $S_I=2 e I f$. After some algebra we obtain an analytic
expression for the Fano factor corresponding to the qubit product
state $|z\rangle$,
\begin{equation}
f_z=\frac{S_I^z}{2eI_z}
=\frac{\left(\Gamma_R^{z}+\Gamma_L^{'z}\right)^2\left((\Gamma_L^{z})^2+(\Gamma_R^{'z})^2\right)-
\left(\Gamma_L^{z}\Gamma_R^{z}+\Gamma_L^{'z}\Gamma_R^{'z}\right)^2 +
4 (\Gamma_L^{z})^2(\Gamma_R^{'z})^2 }
{\left(\Gamma_L^{z}\Gamma_L^{'z}+2\Gamma_L^{z}\Gamma_R^{'z}+\Gamma_R^{z}\Gamma_R^{'z}\right)^2}
.
\end{equation}

We now define the time needed to distinguish between two different
product states $|z_1\rangle$ and $|z_2\rangle$ as\cite{makhlin01}
\begin{equation}
t_{ms}^{z_1z_2}=\left(\frac{\sqrt{S_I^{z_1}}+\sqrt{S_I^{z_2}}}{I_{z_1}-I_{z_2}}\right)^2
, \label{measurement_time_def}
\end{equation}
after which the two corresponding counting distributions are
distinguishable. Distinguishable implies that the overlap of the
counting distributions fulfills\cite{AverinSukhorukovPRL2005}
\begin{equation}
\sum_m \sqrt{P_{z_1}(m,t^{z_1z_2}_{ms})P_{z_2}(m,t^{z_1z_2}_{ms})}
\leq \frac{1}{e}, \label{ProbDistOverlapEq}
\end{equation}
where $P_{z}(m,t)$ corresponds to the qubits' state $|z\rangle$.

According to fundamental principles of quantum mechanics, two
different states cannot be distinguished classically before the
quantum coherence between these states has decayed.\cite{Braginsky}
With our definitions this inequality reads $\Gamma_\varphi^{z_1z_2}
t_{ms}^{z_1z_2} \geq 1$, leading to the definition of the quantum
efficiency
\begin{equation}
\eta^{z_1z_2}=\frac{1}{\Gamma_\varphi^{z_1z_2} t_{ms}^{z_1z_2}} \leq
1,
\end{equation}
for distinguishing the two states $|z_1\rangle$ and $|z_2\rangle$,
using our suggested measurement scheme. One may note that other
definitions of distinguishability, and thus measurement time, gives
other numerical values for the limit of the quantum efficiency. This
is the reason why the quantum limit sometimes in literature is given
as 1/2. The foundation is now established for discussing the
properties of the suggested readout scheme in some analytically
tractable, and relevant cases.

\end{widetext}

\subsection{A single junction detecting a single qubit}
For comparison we may check the limit of a single QPC measuring
the state of a single qubit. This situation is obtained from our
system if the right qubit is uncoupled ($\delta\Gamma_R=0$), and
if the escape rate through the right junction is much larger than
the in-rates $\Gamma_R \gg \Gamma_L^{\pm}$. To leading (zeroth)
order in $\Gamma_L^\pm/\Gamma_R$ the island is always empty, i.e.
$P_{zz}^a=1$, the average current is $I_z= 2 e \Gamma_L^z$, and
the Fano factor is one. The time needed to separate the product
state which differ in the state of the left qubit is obtained from
Eq.~(\ref{measurement_time_def}),
\begin{eqnarray}
& &t_{ms}^{AC}=t_{ms}^{AD}=t_{ms}^{BC}=t_{ms}^{BD}= \\
&=&
 \frac{1}{2}
\frac{1}{\Gamma_L-\sqrt{\Gamma_L^2-\delta\Gamma_L^2}} \approx
\frac{\Gamma_L}{\delta\Gamma_L^2}
\left[1+O\left(\frac{\delta\Gamma_L^2}{\Gamma_L^2}\right)\right],
\nonumber
\end{eqnarray}
where the last equality is relevant in the weak coupling regime
($\delta\Gamma_L\ll\Gamma_L$). The corresponding dephasing rates
are given by Eq.~(\ref{TotalDephasingRate}),
\begin{eqnarray}
\Gamma_\varphi^{AC}&=&\Gamma_\varphi^{AD}=\Gamma_\varphi^{BC}=
\Gamma_\varphi^{BD}= \\
&=& 2\left(\Gamma_L-\sqrt{\Gamma_L^2-\delta\Gamma_L^2}\right)
\approx \frac{\delta\Gamma_L^2}{\Gamma_L}
\left[1+O\left(\frac{\delta\Gamma_L^2}{\Gamma_L^2}\right)\right].
\nonumber
\end{eqnarray}
Thus we find that this readout is indeed quantum limited
$\eta^{AC}=\eta^{BC}=\eta^{CD}=\eta^{BD}=1$, which was first
discussed in Ref.~\onlinecite{gurvitz97}.

\subsection{Negligible Coulomb energy}
Now, going back to the case of two qubits detected by two QPCs, a
limit which is easy to analyze is when the rates do not depend on
whether the doubly occupied state is involved or not, i.e.
$\Gamma_{L/R}^{'z}=\Gamma_{L/R}^{z}$. This regime is e. g. obtained
if the Coulomb energy can be neglected $U=0$. The average current is
then
\begin{equation}
I_{z}=2e\frac{\Gamma_R^{z}\Gamma_L^{z}}
{\Gamma_L^{z}+\Gamma_R^{z}}, \label{non_interacting_current}
\end{equation}
and the noise is the same as for a non-interacting quantum
dot\cite{blanter00}
\begin{equation}
f= \frac{(\Gamma_L^z)^2+(\Gamma_R^z)^2}
{\left(\Gamma_L^z+\Gamma_R^z\right)^2} .
\label{non_interacting_Fano}
\end{equation}
The occupation probabilities of the island are
\begin{eqnarray}
P_{zz}^{a} &=& \frac{(\Gamma_R^{z})^2}
{\left(\Gamma_L^{z}+\Gamma_R^{z}\right)^2}, \qquad P_{zz}^{b} =
\frac{2\Gamma_L^{z}\Gamma_R^{z}}
{\left(\Gamma_L^{z}+\Gamma_R^{z}\right)^2}, \nonumber \\
P_{zz}^{c} & = & \frac{(\Gamma_L^{z})^2}
{\left(\Gamma_L^{z}+\Gamma_R^{z}\right)^2} .
\end{eqnarray}
Now, looking at the symmetric situation where the left and right
tunnelling rates are equal ($\Gamma_L=\Gamma_R=\Gamma$) and the
qubits are equally strongly coupled to the SET
($\delta\Gamma_L=\delta\Gamma_L=\delta\Gamma$), we have the
following expressions for the time needed to distinguish the
different qubit product states,
\begin{eqnarray}
t_{ms}^{AD} &=& \frac{\Gamma}{\delta\Gamma},\qquad
t_{ms}^{BC}=\infty,
\nonumber \\
t_{ms}^{AB} &=& t_{ms}^{AC}=t_{ms}^{BD}=t_{ms}^{CD} =
4\frac{\Gamma}{\delta\Gamma},
\label{non_interacting_measurement_times}
\end{eqnarray}
to lowest order in $\delta\Gamma/\Gamma$. The states $|A\rangle =
|\downarrow\downarrow\rangle$ and $|D\rangle =
|\uparrow\uparrow\rangle$ have the largest difference in average
current, and are the fastest to distinguish. The states $|B\rangle
= |\downarrow\uparrow\rangle$ and $|C\rangle =
|\uparrow\downarrow\rangle$ have the same average current, and
thus we cannot tell them apart in this measurement. To distinguish
all product states one has to measure for at least as long as the
longest measurement time $t_{ms}^{AB}$. The corresponding quantum
efficiencies are
\begin{eqnarray}
\eta^{AD} &=& 1,\qquad \eta^{BC}=0,
\nonumber \\
\eta^{AB}&=&\eta^{AC}=\eta^{BD}=\eta^{CD}=\frac{1}{2},
\label{non_interacting_quatum_efficiencies}
\end{eqnarray}
again to leading order in $\delta\Gamma/\Gamma$. Here we may note
that the quantum efficiency is unity only for telling the states
$|A\rangle$ and $|D\rangle$ apart. The states $|B\rangle$ and
$|C\rangle$ we can not tell apart, but they are still dephased,
giving a zero quantum efficiency.

For the other pairs of states, we find a quantum efficiency of one
half, indicating that the qubits are getting entangled with a degree
of freedom which is not measured. This degree of freedom is the
island charge state, and how the quantum efficiency can be improved
by also detecting this will be discussed in
Sec.~\ref{IslandStateMeasSec}.

In Refs.~\onlinecite{RuskovPRB2003, MaoPRL2004, TrauzettelPRB2006},
measurements which are unable to discern certain product states are
used to create entanglement. Our measurement setup can {\em not} be
used for entangling the qubits, since it always projects the qubits
on a product state. As will be shown in
Sec.~\ref{IslandStateMeasSec}, the inability to separate the
$|B\rangle$ and $|C\rangle$ states is not inherent to the coupling
between the qubits and the measurement device, but rather a
consequence of measuring only the current.

\subsection{Large Coulomb energy}
A maybe more realistic situation is when the Coulomb charging
energy $U$ of the island is large, and especially larger than the
applied drain-source voltage. The in-rates $\Gamma_L^{'z}$ to the
doubly occupied state are then zero for all qubit-states
$|z\rangle$, and the doubly occupied state of the SET island will
be unoccupied. In this situation the steady state island
probabilities are
\begin{equation}
P_{zz}^{a} = \frac{\Gamma_R^{z}} {2\Gamma_L^{z}+\Gamma_R^{z}},
\qquad P_{zz}^{b} = \frac{2\Gamma_L^{z}}
{2\Gamma_L^{z}+\Gamma_R^{z}}, \qquad P_{zz}^{c} =  0 ,
\end{equation}
resulting in the average steady-state current
\begin{equation}
I_{z}=2e\frac{\Gamma_R^{z}\Gamma_L^{z}}
{2\Gamma_L^{z}+\Gamma_R^{z}},
\end{equation}
and Fano factor\cite{nazarov96}
\begin{equation}
f= \frac{4(\Gamma_L^z)^2+(\Gamma_R^z)^2}
{\left(2\Gamma_L^z+\Gamma_R^z\right)^2} . \label{FanoLargeCoulomb}
\end{equation}
We may note that the expressions for the current and Fano factor
can formally be obtained from Eqs.~(\ref{non_interacting_current})
and (\ref{non_interacting_Fano}) by letting $\Gamma_R^z
\rightarrow \Gamma_R^z/2$. Again looking at the symmetric
situation where the left and right tunnelling rates are equal
($\Gamma_L=\Gamma_R=\Gamma$), and the qubits are equally strongly
coupled to the SET ($\delta\Gamma_L=\delta\Gamma_L=\delta\Gamma$),
we have the following expressions for the time needed to
distinguish the different qubit product states,
\begin{eqnarray}
t_{ms}^{AD} &=& \frac{5}{3}\frac{\Gamma}{\delta\Gamma},\qquad
t_{ms}^{AB}=t_{ms}^{CD} = \frac{15}{4}\frac{\Gamma}{\delta\Gamma},
\nonumber \\
t_{ms}^{AC}=t_{ms}^{BC}=t_{ms}^{BD} &=&
15\frac{\Gamma}{\delta\Gamma},
\end{eqnarray}
where we again assumed weak coupling to the qubits $\delta\Gamma
\ll \Gamma$, and presented the results to leading order in
$\delta\Gamma/\Gamma$. The states $|A\rangle =
|\downarrow\downarrow\rangle$ and $|D\rangle =
|\uparrow\uparrow\rangle$ again have the largest difference in
average current, and are the fastest to distinguish. The states
$|B\rangle = |\downarrow\uparrow\rangle$ and $|C\rangle =
|\uparrow\downarrow\rangle$ no longer give the same average
current, and thus we can tell them apart. To distinguish all
product states one has to measure for at least as long as the
longest measurement time $t_{ms}^{AB}$. The corresponding quantum
efficiencies are
\begin{eqnarray}
\eta^{AD} &=& \frac{9}{10},\qquad \eta^{AB}=\eta^{CD} =
\frac{4}{5},
\nonumber \\
\eta^{AC}=\eta^{BD}&=&\frac{1}{5}, \qquad \eta^{BC} =
\frac{1}{10}, \label{LargeCooulombEnergyQuantumEfficiencies}
\end{eqnarray}
to leading order in $\delta\Gamma/\Gamma$. Here all quantum
efficiencies are finite, since we can distinguish all product
states. Also, the quantum efficiency for separating the $|A\rangle$
and $|D\rangle$ state is lowered to $9/10$ compared to unity in the
noninteracting case, which is due to the differences in island state
probabilities between the $|A\rangle$ and $|D\rangle$ state. In
fact, by choosing $\Gamma_L=\Gamma, \Gamma_R=2\Gamma,
\delta\Gamma_L=\delta\Gamma$ and $\delta\Gamma_R=2\delta\Gamma$ the
quantum efficiency $\eta^{AD}$ is again unity, and the measurement
times and quantum efficiencies are given by the expressions for the
noninteracting case in
Eqs.~(\ref{non_interacting_measurement_times}) and
(\ref{non_interacting_quatum_efficiencies}).

\section{Measuring the island state}
\label{IslandStateMeasSec} When the tunneling rates are slow enough,
one may perform a time-resolved measurement of the island charge
state, e.g. using another RF-SET\cite{LuNature2003}. Plotting the
island charge as a function of time, each tunnel event appears as an
almost vertical line, and from this information it is
straightforward to determine the full counting statistics of the
electrons passing the detector, as shown by Gustavsson {\it et al}.
\cite{gustavsson06}

\subsection{Measuring the full time-trace}
The time the system waits in island state $a$ ($t_w^a$) and $c$
 ($t_w^c$) are random variables, exponentially distributed with parameters $\Gamma_L$ and
$\Gamma_R'$, respectively (neglecting the coupling to the qubits).
>From state $b$, the system escapes both to $a$ and $c$, so this
waiting time is exponentially distributed with parameter
$\Gamma_L'+\Gamma_R$. By labeling the $b$ state waiting times also
with the destination state ($a$ or $c$), one gets two separate
waiting times ($t_w^{ba}$ and $t_w^{bc}$), each exponentially
distributed, with parameter $\Gamma_R$ and $\Gamma_L'$ respectively.
The average waiting time equals the inverse rate, e. g. $\langle
t_w^a \rangle = \Gamma_L^{-1}$, and the variance is decreasing as
$\Gamma_L^{-2} n^{-1}$, where $n$ is the number of tunnel events
$(a\rightarrow b)$ where $t_w^a$ has been sampled. The average
number of tunnel events increases linearly with time, $n_a=t P^a 2
\Gamma_L$, $n_{ba}=t P^b \Gamma_R$, $n_{bc}=t P^b \Gamma_L'$ and
$n_c=t P^c 2 \Gamma_R'$.

Now, including the coupling to the qubits, each of the four rates
may have one of two possible values $\Gamma_{L/R}^{(')+}$ or
$\Gamma_{L/R}^{(')-}$. The combined probability distribution for the
four different waiting times will now depend on the qubits' state.
We again use Eq.~(\ref{ProbDistOverlapEq}) as the definition of the
measurement time, with the difference that the probability
distribution now is a function of four continuous random variables.

In the weak coupling regime $\delta\Gamma \ll \Gamma$ the individual
distributions approach a Gaussian form, in the relevant regime of
all $n_a,n_{ba},n_{bc}, n_c \gg 1$. The overlap for the $t_w^a$
distributions is
\begin{equation}
\int_0^{\infty} dt_w \sqrt{P_{z_1}(t_w,t)P_{z_2}(t_w,t)} = e^{- n_a
\delta\Gamma_L^2/2\Gamma_L^2},
\end{equation}
where $n_a$ is the average number of $a\rightarrow b$ tunnel events,
neglecting the coupling to the qubits. This expression is valid to
lowest non-vanishing order in all $\delta\Gamma/\Gamma$. Since the
waiting times are independently distributed random variables we get
the total overlap as a product of the four individual overlaps,
giving the measurement time
\begin{eqnarray}
 \left[t_{ms}^{z_1z_2}\right]^{-1} &=&
\Theta_L(z_1,z_2) \left[ P_a \frac{\delta\Gamma^2_L}{\Gamma_L} +
\frac{P_b}{2}
\frac{(\delta\Gamma_L')^2}{\Gamma_L'}\right] + \nonumber  \\
& +& \Theta_R(z_1,z_2) \left[ \frac{P_b}{2}
\frac{\delta\Gamma^2_R}{\Gamma_R}+ P_c
\frac{(\delta\Gamma_R')^2}{\Gamma_R'}\right],
\label{RateMeasurementTime}
\end{eqnarray}
where $\Theta_{L/R}(z_1,z_2)=1$ if the states $|z_1\rangle$ and
$|z_2\rangle$ differ for the left/right qubit, and
$\Theta_{L/R}(z_1,z_2)=0$ otherwise. Comparing with the weak
coupling expression of the total dephasing rate in
Eq.~(\ref{TotalDephasingRate}), we find the quantum efficiency of
recording all tunnel events to be unity, to lowest order in
$\delta\Gamma/\Gamma$. Thus, during the measurement, all the
information about the qubits' state which is extracted is
transferred into information about the four different tunneling
rates.

\section{Quantum efficiency of current measurements revisited}
\label{CurrentMeasurementsRevisitedSection} 
\begin{figure}[!ht]
\includegraphics[width=9cm]{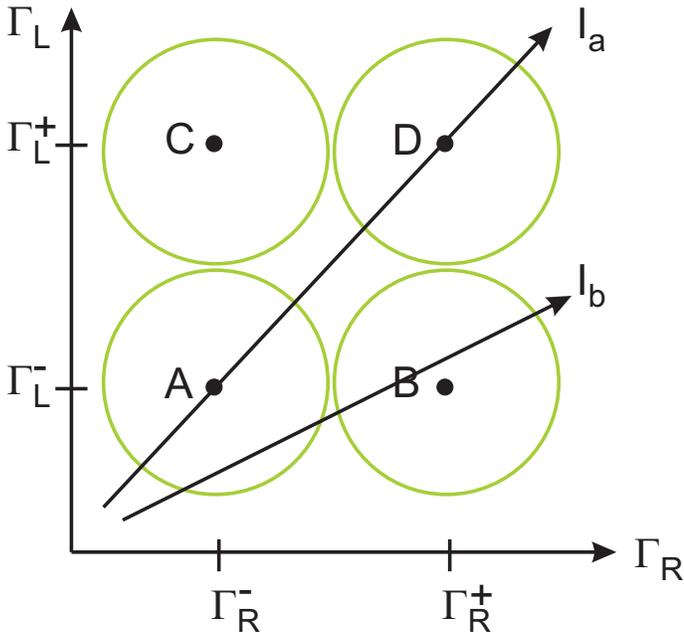}
\caption{A schematic representation of the different two-dimensional
rate distributions corresponding to the 4 different qubit product
states A-D. A current measurement is equivalent to projecting the
distributions on the axis $I_a$ in the case of negligible Coulomb
energy, and on $I_b$ in the case of large Coulomb
energy.}\label{CurrentVsRatesFig}
\end{figure}

Knowing that the full
information is in the distribution of the rates, we can now
reexamine the current measurement in terms of how much information
it gives about the rates. In general the current is a function of
all four rates, and the analysis gets complicated. For simplicity we
here only reanalyze the two cases we analyzed analytically above:
negligible and large Coulomb energy. In both these cases there is
only one independent non-zero rate per junction. Thus the current is
a function of only two independent random variables, and we may
straightforwardly visualize the situation. In
Fig.~\ref{CurrentVsRatesFig} the pair of rates corresponding to the
four different qubit product states A-D are shown as dots in a
two-dimensional rate diagram.
In the weak coupling regime the distributions are the bivariate
Gaussians, centered around the corresponding state. The width of
the distributions decrease with time as $~1/\sqrt{t}$. The green
circles indicates where the distributions decreased to $1/e$, at
the measurement time
$t_{ms}^{AB}=t_{ms}^{AC}=t_{ms}^{BD}=t_{ms}^{CD}$. Since the
distance between A and D is $\sqrt{2}$ longer, these states
separate earlier, and as is shown by
Eq.~(\ref{RateMeasurementTime}) this measurement time is only
half, $t_{ms}^{AD}=t_{ms}^{AB}/2$.

The one-dimensional distribution of the current is given by
integrating the two-dimensional rate-distribution along the axis
perpendicular to the vector \cite{GrimmettStirzaker}
\begin{equation}
\left[ \partial_{\Gamma_L} I(\Gamma_L,\Gamma_R),
\partial_{\Gamma_R} I(\Gamma_L,\Gamma_R) \right] .
\end{equation}

In the case of negligible Coulomb energy, the current formula is
symmetric in the two rates, and the projection axis is diagonal, as
shown by the $I_a$-axis. It is obvious that the separation of the
states A and D has full quantum efficiency, since the line
connecting the two states fall on the current axis. In general the
quantum efficiency is given by
$\eta^{z_1z_2}=\cos^2{\alpha^{z_1z_2}_I}$, where $\alpha^{z_1z_2}_I$
is the angle between the current axis and the line connecting the
states $|z_1\rangle$ and $|z_2\rangle$. Since the pairs of states
AB, AC, BD, and CD are connected by horizontal or vertical lines,
the corresponding quantum efficiency is $\cos^2{(\pi/4)}=1/2$.
Furthermore, the line connecting the B and C states is orthogonal to
the current axis, giving a
 zero ($\cos^2{(\pi/2)}=0$) quantum efficiency, i.e. the projection of
the two distributions always overlap completely. This simple
geometrical interpretation is in full agreement with
Eq.~(\ref{non_interacting_quatum_efficiencies}).

In the other analytically tractable case of large Coulomb
interaction, the current axis $I_b$ forms an angle
$\gamma=\arctan{(1/2)}$ with the horizontal $\Gamma_R$-axis. Noting
that $\cos^2{\gamma}=4/5$, and that $\cos^2{(\gamma-\pi/4)}=9/10$ we
see that the geometrical interpretation also reproduces
Eq.~(\ref{LargeCooulombEnergyQuantumEfficiencies}).

\subsection{Measuring average island charge}
Measuring the full time-trace of all tunnel events has so far only
been demonstrated for tunnel rates on the order of kHz
\cite{gustavsson06}. For faster readout schemes a more realistic
setup would be to measure the average island charge $\langle n
\rangle$, in addition to the average current through the device.
The time-dependent distribution of the average charge $\langle n
\rangle$ can also be analyzed in terms of the rate distributions.
One may note that the vector corresponding to the average charge
measurement is perpendicular to the current measurement vector in
the symmetric setup, i.e. $\Gamma_L=\Gamma_R$ for negligible
Coulomb energy, and $\Gamma_R=2\Gamma_L$ for large Coulomb energy.
In this case the distribution of average charge is independent of
the current distribution, and we recover full quantum efficiency
in separating all product states, even if the full time-trace
cannot be recorded.

\section{Conclusion and Discussion}
\label{DiscussionSection} We have described how to read out two
charge qubits using a single SET. After deriving general kinetic
equations for the whole system, we focussed on the properties of
single-shot readout. Any interdot coupling during the measurement
will reduce the fidelity, so we considered the case when the
inter-dot qubit tunneling is turned off during measurement
($\Delta_L=\Delta_R=0$). Detecting the charge of the SET island in
real-time corresponds to measuring the individual tunnel rates, and
is always quantum efficient. The current is a function of the
tunneling rates, and the quantum efficiency in this case depends on
which qubit states should be separated, as well as the biasing
conditions of the SET. In some symmetric cases, a combined
measurement of average current and average island charge gives full
quantum efficiency for separating all four product states. The setup
can {\em not} be used for entangling the qubits, since the
measurement always projects onto a product state.

The possibility to have one readout device for two qubits can be
used to optimize the design of many-qubit circuits. One drawback is
that the qubits always have to be read out simultaneously. This may
not be a severe problem, since in many quantum algorithms the
readout is performed at the end. Also, there is an obvious advantage
in reducing the number of readout lines attached to your circuit.
This has to be weighed against the reduced signal-to-noise ratio in
the readout, coming from the fact that four different states should
be separated, instead of two.

\section{Acknowledgements}
This work was supported by the European Commission through the
IST-015708 EuroSQIP integrated project.

\end{document}